\documentclass[twocolumn,aps,prx,showpacs,nofootinbib,nobibnotes]{revtex4}%
\usepackage{amsmath}
\usepackage{amsfonts}
\usepackage{mathrsfs}
\usepackage[colorlinks,linkcolor=blue,anchorcolor=blue,citecolor=blue,urlcolor=black]%
{hyperref}
\usepackage{graphicx}
\usepackage{dcolumn}
\usepackage{bm}
\usepackage{epsfig}
\usepackage{amssymb}%
\def\be{\begin{equation}}
\def\ee{\end{equation}}
\begin{document}
\title{Cooling mechanical resonators to quantum ground state from room temperature}
\author{Yong-Chun Liu$^{1,2}$, Rui-Shan Liu$^{1}$, Chun-Hua Dong$^{3}$, Yan Li$^{1,2}%
$, Qihuang Gong$^{1,2}$, and Yun-Feng Xiao$^{1,2}$}
\email{yfxiao@pku.edu.cn}
\altaffiliation{URL: www.phy.pku.edu.cn/$\sim$yfxiao/index.html}

\affiliation{$^{1}$State Key Laboratory for Mesoscopic Physics and School of Physics,
Peking University, Beijing 100871, P. R. China}
\affiliation{$^{2}$Collaborative Innovation Center of Quantum Matter, Beijing 100871,
People's Republic of China}
\affiliation{$^{3}$Key Laboratory of Quantum Information, University of Science and
Technology of China, Hefei 230026, People's Republic of China}
\date{\today}

\begin{abstract}
Ground-state cooling of mesoscopic mechanical resonators is a fundamental
requirement for test of quantum theory and for implementation of quantum
information. We analyze the cavity optomechanical cooling limits in the
intermediate coupling regime, where the light-enhanced optomechanical coupling
strength is comparable with the cavity decay rate. It is found that in this
regime the cooling breaks through the limits in both the strong and weak
coupling regimes. The lowest cooling limit is derived analytically at the
optimal conditions of cavity decay rate and coupling strength. In essence,
cooling to the quantum ground state requires $Q_{\mathrm{m}}>2.4n_{\mathrm{th}%
}$, with $Q_{\mathrm{m}}$ being the mechanical quality factor and
$n_{\mathrm{th}}$ being the thermal phonon number. Remarkably, ground-state
cooling is achievable starting from room temperature, when mechanical
$Q$-frequency product $Q_{\mathrm{m}}{\nu>1.5}\times10^{13}$, and both of the
cavity decay rate and the coupling strength exceed the thermal decoherence
rate. Our study provides a general framework for optimizing the backaction
cooling of mesoscopic mechanical resonators.

\end{abstract}

\maketitle

Cavity optomechanics \cite{RevSci08,RevPhy09,RevRMP13,RevCPB13,Pierre}
provides an important platform for manipulation of mesoscopic mechanical
resonators in the quantum regime. A prominent example is motional ground-state
cooling, which reduces the thermal noise of the mechanical resonator all the
way to the quantum ground state \cite{GSNat11,GSNat11-2}. This offers as the
first crucial step for most applications such as the exploration of
quantum-classical boundary \cite{superPRL11,superPRL12,superPRA13zqyin} and
quantum information processing \cite{qcPRL03,NetworkPRL10,QIPPRL12}. Recently
cooling of mechanical resonators has been demonstrated using various
approaches including pure cryogenic cooling \cite{GSNat10}, feedback cooling
\cite{FBCooPRL98,FBCooPRL99,FBCooNat06,FBCooPRL07,FBCooPRL07-2} and
cavity-assisted backaction cooling
\cite{CooNat06,CooNat06-2,CooPRL06,CooNatPhys08,CooNatPhys09-1,CooNatPhys09-2,CooNatPhys09-3,CooNat10,GSNat11,GSNat11-2,CooPRA11}%
, along with many theoretical and experimental efforts on novel cooling
schemes, such as cooling with dissipative coupling
\cite{DCPRL09,LiPRL2009,DCPRL11,DCPRA13,myyanPRA13}, quadratic coupling
\cite{quadCooPRA10}, single-photon strong coupling \cite{SSCCooPRA12}, hybrid
systems \cite{Atom09PRA,AtomPRA13}, laser pulse modulations
\cite{PulPNAS11,PulPRA11-1,PulPRA11-2,PulPRL11,PulPRL12} and dissipation
modulations \cite{ycliuDC13}. It is theoretically shown that ground-state
cooling is possible in the resolved sideband regime
\cite{PRL07-1,PRL07-2,PRA08}, where the mechanical resonance frequency is
greater than the decay rate of the optical cavity, indicating the resolved
mechanical sideband from cavity mode spectrum. These analyses are in the weak
coupling regime, where the light-enhanced optomechanical coupling strength
${G}$ is weak compared with the cavity decay rate ${\kappa}$, and thus the
coupling is regarded as a perturbation to the optical field. Within this
regime a larger coupling strength is better since the net cooling rate
(optical damping rate) scales as ${\Gamma}_{\mathrm{wk}}={4G}^{2}/\kappa$. On
the other hand, when ${G\gg\kappa}$, the system is in the strong coupling
regime \cite{SCPRL08,SCNJP08,SCNat09,SCNat11,SCNat12,SCNat13,ycliuDC13}, where
normal-mode splitting occurs and the phonon occupancy exhibits Rabi-like
oscillation with reversible energy exchange between optical and mechanical
modes. Then the cooling rate saturates with the maximum value of ${\Gamma
}_{\mathrm{str}}={\kappa}$, and thus a larger cavity decay rate ${\kappa}$ is
better. However, in this case, a large ${\kappa}$ in turn makes the system
away from the strong coupling regime. As a result, the optimal cooling is
expected for the intermediate coupling regime, where the coupling strength
${G}$ is comparable with the cavity decay rate ${\kappa}$.

In the weak coupling regime, the perturbative approach \cite{PRL07-1,PRL07-2}
is widely adopt. In the intermediate and strong coupling regimes, however, the
perturbative approach fails since the optomechanical coupling can no longer be
considered as a perturbation. One way to overcome this problem is to employ
the covariance approach, where all the mean values of the second-order moments
are computed with the linearized optomechanical interaction
\cite{SCNJP08,ycliuDC13}. In this paper, we use this non-perturbative approach
to analyze the optimal cooling limits in the full parameter range and derive
the optimal parameters, including laser detuning, cavity decay rate and
optomechanical coupling strength. We then find that the optimal cooling is
reached with ${G}\sim0.6\kappa$, which is in the intermediate coupling regime.
Finally we show the unique advantage of cooling in this regime, where
room-temperature ground-state cooling is achievable for mechanical
$Q$-frequency product $Q_{\mathrm{m}}{\nu>1.5}\times10^{13}$, which is within reach for current experimental conditions \cite{highQPRL14}.

\begin{figure}[tb]
\centerline{\includegraphics[width=\columnwidth]{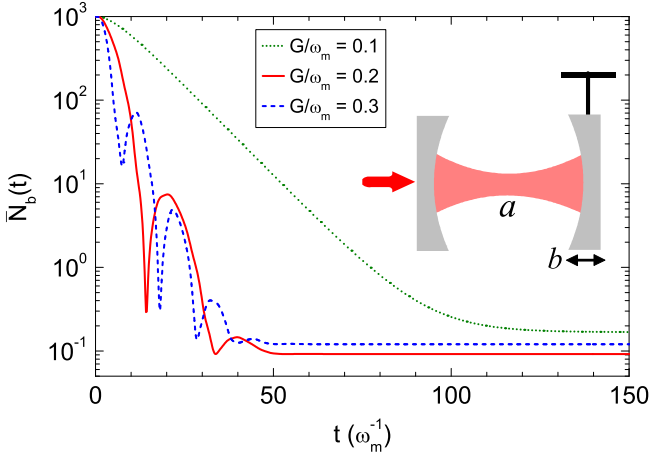}}
\caption{(color online) Time evolution of the mean phonon number $\bar{N}%
_{b}(t)$ for $G/{\omega_{\mathrm{m}}}=0.1$ (green dotted curve), $0.2$ (red
solid curve) and $0.3$ (blue dashed curve). Other parameters are
$\Delta^{\prime}=-{\omega_{\mathrm{m}}}$, $\kappa/{\omega_{\mathrm{m}}}=0.5$,
${\gamma/\omega_{\mathrm{m}}=10}^{-5}$ and $n_{\mathrm{th}}=10^{3}$. Inset:
Sketch of the optomechanical system.}%
\label{Fig1}%
\end{figure}

We consider the general model of an optomechanical system, as shown in the set
of Fig. \ref{Fig1}. A mechanical mode interacts with an optical resonance mode
which is driven by a coherent laser. The system Hamiltonian reads $H={{\omega
}_{\mathrm{c}}a^{\dag}a}+{\omega_{\mathrm{m}}b^{\dag}b+ga^{\dag}a{(b+b^{\dag
})+({\Omega e}}}^{-i\omega t}{a^{\dag}+\Omega}^{\ast}{{{e}}}^{i\omega t}{a)}$.
Here ${\omega}_{\mathrm{c}}$ (${\omega_{\mathrm{m}}}$) is the angular
resonance frequency of the optical mode $a$ (mechanical mode $b$); the third
term describes the optomechanical interaction \cite{LawPRA95}, with $g$ being
the single-photon coupling rate; the last term describes the driving of the
input laser with driving strength ${{\Omega}}$ and frequency $\omega$. The
coherent driving shifts the optical states and thereby shifts the mechanical
states via the optical force. Thus the operators are rewritten as
${a\rightarrow\alpha+a}_{1}$ and $b\rightarrow\beta+b_{1}$, where ${\alpha}$
($\beta$) represents the steady state value of the optical (mechanical) mode,
and ${a}_{1}$ ($b_{1}$) stands for the corresponding fluctuation operator. For
strong intracavity field $\left\vert {\alpha}\right\vert \gg1$, the
Hamiltonian is approximated as quadratic type given by $H_{L}=-\Delta^{\prime
}{a_{1}^{\dag}a}_{1}+{\omega_{\mathrm{m}}{b_{1}^{\dag}b}_{1}+G(a_{1}^{\dag
}+a_{1})(b_{1}+b_{1}^{\dag}})$, where $G=\left\vert {\alpha}\right\vert {g}$
describes the light-enhanced optomechanical coupling strength and ${\Delta
}^{\prime}=\omega-{{\omega}_{\mathrm{c}}}+2{G}^{2}/{\omega_{\mathrm{m}}}$
denotes the optomechanical-coupling modified detuning. In the above derivation
we have absorbed the phase factor of ${\alpha}$ into the operator $a$.

The system evolution is governed by master equation described by $\dot{\rho
}=i[\rho,H_{L}]+\kappa\mathcal{D}[{a}_{1}]\rho+\gamma(n_{\mathrm{th}%
}+1)\mathcal{D}[{{b}_{1}}]\rho+\gamma n_{\mathrm{th}}\mathcal{D}[{{b_{1}%
^{\dag}}}]\rho$. Here $\mathcal{D}[\hat{o}]\rho=\hat{o}\rho\hat{o}^{\dag
}-{({\hat{o}^{\dag}\hat{o}\rho+\rho\hat{o}^{\dag}\hat{o})/2}}$ (${{\hat{o}=a}%
}_{1}$, ${{b}_{1}}$, ${{b_{1}^{\dag}}}$) denotes the Lindblad dissipators;
$\kappa$ ($\gamma$) represents the dissipation rate of the optical cavity
(mechanical) mode; $n_{\mathrm{th}}=1/(e^{\hbar{\omega_{\mathrm{m}}%
/k}_{\mathrm{B}}T}-1)$ corresponds to the bath thermal phonon number at the
environmental temperature $T$. Using the master equation, the mean phonon
number $\bar{N}_{b}=\langle{b_{1}^{\dag}b}_{1}\rangle=\mathrm{Tr}(\rho
{b_{1}^{\dag}b}_{1})$ can be determined by a linear system of ordinary
differential equations involving all the second-order moments
\cite{SCNJP08,ycliuDC13,ycliuSC14}, i. e., $\partial_{t}\langle\hat{o}_{i}%
\hat{o}_{j}\rangle=\mathrm{Tr}(\dot{\rho}\hat{o}_{i}\hat{o}_{j})=\sum
_{k,l}\eta_{k,l}\langle\hat{o}_{k}\hat{o}_{l}\rangle$, where $\hat{o}_{i}$,
$\hat{o}_{j}$, $\hat{o}_{k}$ and $\hat{o}_{l}$ are one of the operators ${{a}%
}_{1}$, ${{b}_{1}}$, ${{a_{1}^{\dag}}}$ and ${{b_{1}^{\dag}}}$. Initially, the
mean phonon number is equal to the bath thermal phonon number, i. e., $\bar
{N}_{b}(t=0)=n_{\mathrm{th}}$, and other second-order moments are zero.

In Fig. \ref{Fig1}, we plot the exact numerical results of typical time
evolution of the mean phonon number $\bar{N}_{b}(t)$ for various coupling
strength $G/{\omega_{\mathrm{m}}}=0.1$, $0.2$ and $0.3$ with the given cavity
decay rate $\kappa/{\omega_{\mathrm{m}}}=0.5$. It can be found that, for
$G={0.2}\kappa$, the mean phonon number decays monotonically, corresponding to
the weak coupling regime. As the coupling strength increases to $G={0.4}%
\kappa$, non-monotonicity appears, which reveals that the system reaches the
intermediate coupling regime, with a lower steady-state cooling limit. For
stronger coupling $G={0.6}\kappa$, the oscillations become more notable.
However, the cooling limit is higher than that for $G={0.4}\kappa$, which is a
result of the stronger quantum backaction.

To shed light on the lower cooling limit in the intermediate coupling regime,
we calculate the steady-state cooling limit in the full parameter range. By
applying the Routh-Hurwitz criterion \cite{StablePRA87}, it is found that the
system reaches a steady state with the stability condition given by
\begin{subequations}
\begin{align}
\Delta^{\prime}  &  <0,\label{Stb1}\\
G^{2}  &  <\frac{(4\Delta^{\prime2}+{\kappa}^{2}){\omega_{\mathrm{m}}}}%
{{16}\left\vert \Delta^{\prime}\right\vert }. \label{Stb2}%
\end{align}
Here Eq. (\ref{Stb1}) implies the red detuning laser input, and Eq.
(\ref{Stb2}) shows that the coupling strength cannot be too strong. Under this
condition, when the system reaches the steady state, the derivatives
$\partial_{t}\langle\hat{o}_{i}\hat{o}_{j}\rangle$ all become zero, and thus
the exact solutions for the steady-state cooling limits can be obtained by
solving the algebraic equations $\mathrm{Tr}(\dot{\rho}\hat{o}_{i}\hat{o}%
_{j})=0$. The cooling limits can concisely be written as $n_{\mathrm{s}%
}=An_{\mathrm{th}}+B$, where $A$ and $B\,$are expressions determined by the
parameters $\Delta^{\prime}$, ${\omega_{\mathrm{m}}}$, $G$, ${\kappa}$ and
${\gamma}$. To provide more physical insights, we divide the steady-state
cooling limits into two parts
\end{subequations}
\begin{equation}
n_{\mathrm{s}}=n_{\mathrm{s}}^{\mathrm{(1)}}+n_{\mathrm{s}}^{\mathrm{(0)}}.
\end{equation}
Here $n_{\mathrm{s}}^{\mathrm{(1)}}=An_{\mathrm{th}}$ describes the classical
cooling limit, which originates from the mechanical dissipation and is
proportional to the environmental thermal phonon number $n_{\mathrm{th}}%
$;\emph{ }$n_{\mathrm{s}}^{\mathrm{(0)}}=B$ denotes the quantum cooling limit
which originates from the quantum backaction and does not depend on
$n_{\mathrm{th}}$.

\begin{figure*}[tb]
\centerline{\includegraphics[width=18cm]{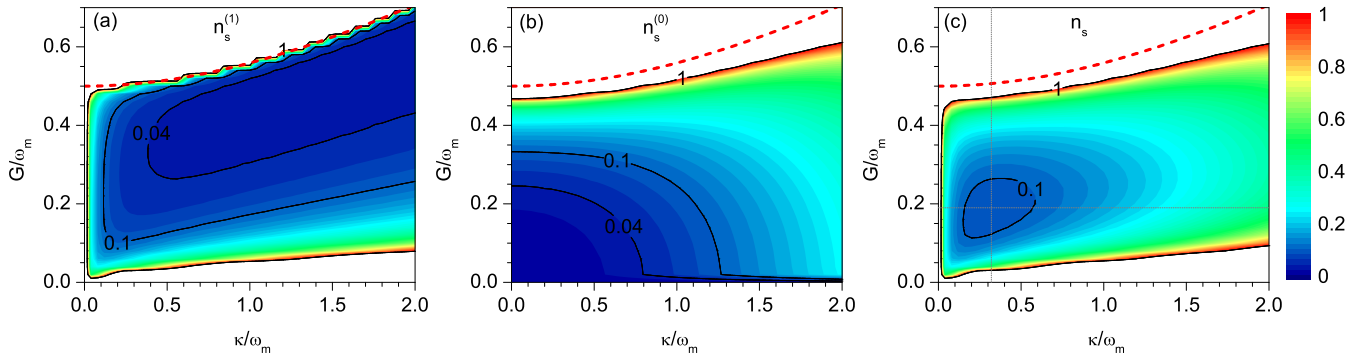}}
\caption{(color online) Steady-state classical cooling limit $n_{\mathrm{s}%
}^{\mathrm{(1)}}$ (panel a), quantum cooling limit $n_{\mathrm{s}%
}^{\mathrm{(0)}}$ (panel b) and total cooling limit $n_{\mathrm{s}}$ (panel c)
as functions of\ the cavity decay rate ${\kappa}$ and coupling strength $G$.
The red dashed curves correspond to the boundary of the stability condition
given by Eq. (\ref{Stb2}). In (c), the vertical and horizontal lines denotes
the optimal ${\kappa}$ and $G$ given by Eqs. (\ref{ks}) and (\ref{Gs}). Other
parameters: $\Delta^{\prime}=-{\omega_{\mathrm{m}}}$, ${\gamma/\omega
_{\mathrm{m}}=10}^{-5}$ and $n_{\mathrm{th}}=10^{3}$.}%
\label{Fig2}%
\end{figure*}

\begin{figure}[tb]
\centerline{\includegraphics[width=\columnwidth]{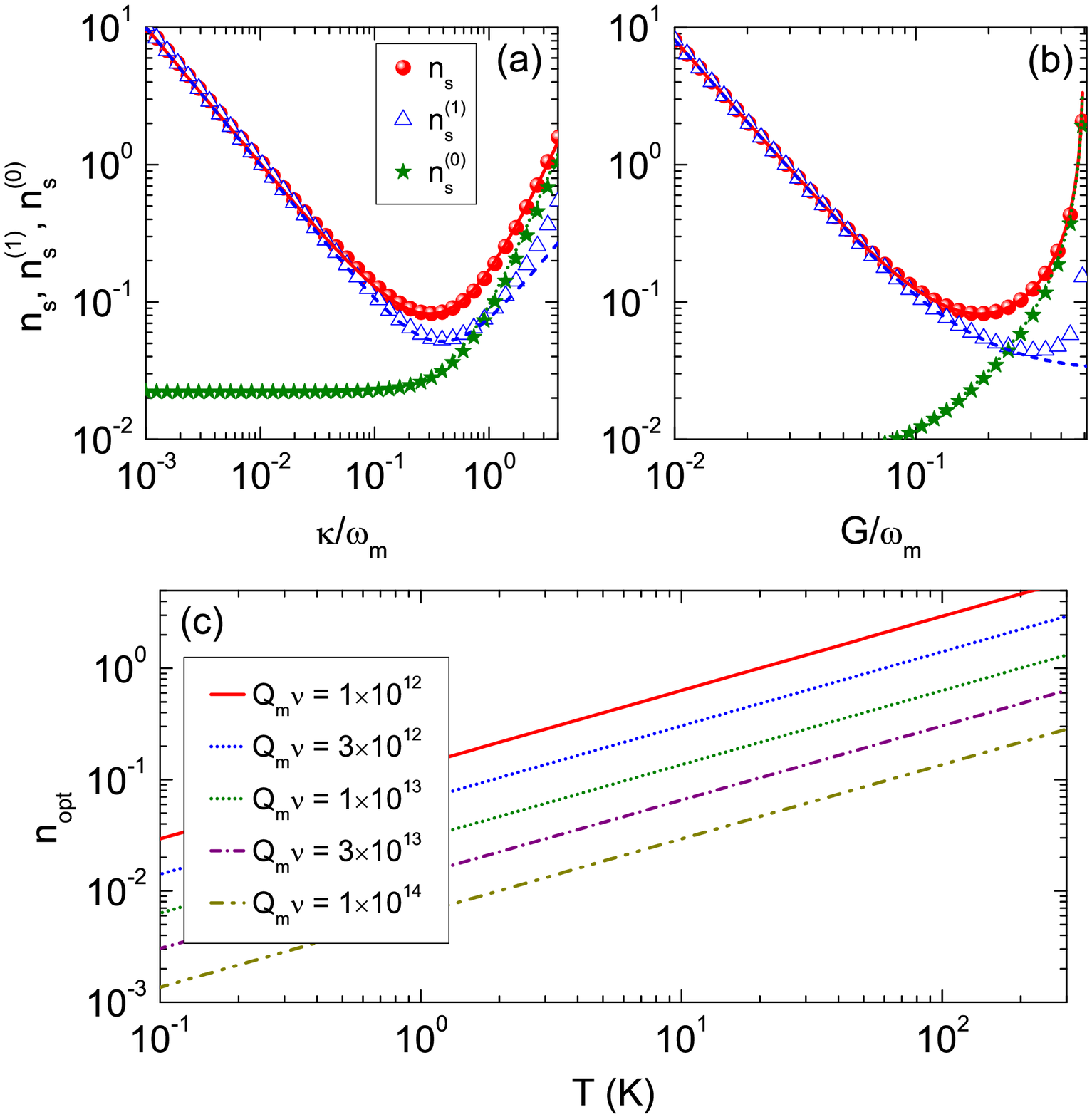}}
\caption{(color online) Cooling limits $n_{\mathrm{s}}$ (red circles and red
solid curves), $n_{\mathrm{s}}^{\mathrm{(1)}}$ (blue triangles and blue dashed
curves) and $n_{\mathrm{s}}^{\mathrm{(0)}}$ (green stars and green dotted
curves) as functions of ${\kappa}$ for $G={G}_{\mathrm{opt}}$ (a) and as
functions of $G$ for $\kappa={\kappa}_{\mathrm{opt}}$ (b). The circles,
triangles and stars are the numerical results, and the curves are the
analytical results given by Eqs. (\ref{nscR}) and (\ref{nsqR}). Other
parameters: $\Delta^{\prime}=-{\omega_{\mathrm{m}}}$, ${\gamma/\omega
_{\mathrm{m}}=10}^{-5}$ and $n_{\mathrm{th}}=10^{3}$. (c) $n_{\mathrm{opt}}$
as a function of the environmental temperature $T$ for $Q_{\mathrm{m}}{\nu
=10}^{12}$ (red solid curve), $3\times{10}^{12}$ (blue dashed curve),
${10}^{13}$ (green dotted curve), $3\times{10}^{13}$ (purple dash-dotted
curve) and ${10}^{14}$ (orange dash-dot-dotted curve).}%
\label{Fig3}%
\end{figure}

In the unresolved sideband regime (${\kappa\gg\omega_{\mathrm{m}}}$) where the
mechanical sideband cannot be resolved from cavity mode spectrum, the optimal
quantum cooling limit is obtained at the detuning $\Delta^{\prime}=-\kappa/2$
with $n_{\mathrm{s}}^{\mathrm{(0)}}\simeq\kappa/(4{\omega_{\mathrm{m}}})\gg1$,
which prevents ground-state cooling \cite{PRL07-1,PRL07-2}. Thus, in the
following we focus on the resolved sideband regime (${\omega_{\mathrm{m}}%
\gg\kappa}$). In this case the optimal detunings for both the classical and
quantum cooling limits are near $\Delta^{\prime}=-{\omega_{\mathrm{m}}}$,
where the rotating-wave interaction characterized by the term $G(a_{1}^{\dag
}b_{1}+a_{1}b_{1}^{\dag})$ is on resonant, leading to the maximum energy
transfer from the mechanical mode to the anti-Stokes sideband. Meanwhile, the
counter-rotating-wave interaction $G(a_{1}^{\dag}b_{1}^{\dag}+a_{1}b_{1})$ is
off resonant, which has minor contribution to the heating process. Under the
condition ${\omega_{\mathrm{m}}\gg(\kappa,G)\gg\gamma}$ and $\Delta^{\prime
}=-{\omega_{\mathrm{m}}}$, we obtain approximate analytical expression of the
cooling limits as
\begin{subequations}
\begin{align}
n_{\mathrm{s}}^{\mathrm{(1)}}\left\vert _{\Delta^{\prime}=-{\omega
_{\mathrm{m}}}}\right.   &  \simeq\frac{{4G}^{2}+{\kappa}^{2}}{{4G}^{2}%
{\kappa}}{\gamma}n_{\mathrm{th}},\label{nscR}\\
n_{\mathrm{s}}^{\mathrm{(0)}}\left\vert _{\Delta^{\prime}=-{\omega
_{\mathrm{m}}}}\right.   &  \simeq\frac{{\kappa}^{2}{+}8{G}^{2}}%
{{16(\omega_{\mathrm{m}}^{2}-4G}^{2})}, \label{nsqR}%
\end{align}
These limits are valid in the weak, intermediate and strong coupling regimes.
In particular, in the weak coupling regime (${\kappa\gg G}$), the cooling
limits reduce to $n_{\mathrm{s}}^{\mathrm{(1)}}\simeq n_{\mathrm{th}}%
{\gamma\kappa}/({4G}^{2})$ and $n_{\mathrm{s}}^{\mathrm{(0)}}\simeq{\kappa
}^{2}/{(16\omega_{\mathrm{m}}^{2}})$, which agree with the perturbation
approaches \cite{PRL07-1,PRL07-2}. For strong coupling regime (${G\gg\kappa}%
$), the cooling limits are simplified as $n_{\mathrm{s}}^{\mathrm{(1)}}\simeq
n_{\mathrm{th}}{\gamma}/{\kappa}$ and $n_{\mathrm{s}}^{\mathrm{(0)}}\simeq
{G}^{2}/[2{(\omega_{\mathrm{m}}^{2}-4G}^{2})]$.

In Fig. \ref{Fig2} we plot the exact numerical results of the cooling limits
$n_{\mathrm{s}}^{\mathrm{(1)}}$, $n_{\mathrm{s}}^{\mathrm{(0)}}$ and
$n_{\mathrm{s}}$ as functions of ${\kappa}$ and $G$ for $\Delta^{\prime
}=-{\omega_{\mathrm{m}}}$, ${\gamma/\omega_{\mathrm{m}}=10}^{-5}$ and
$n_{\mathrm{th}}=10^{3}$. For the classical cooling limit $n_{\mathrm{s}%
}^{\mathrm{(1)}}$, within the stable region, a larger ${G}$ and a larger
$\kappa$ lead to a lower cooling limit, as shown in Fig. \ref{Fig2}(a). Note
that the classical cooling limit can be expressed as $n_{\mathrm{s}%
}^{\mathrm{(1)}}=n_{\mathrm{th}}{\gamma/\Gamma}$, where ${\Gamma}$ is the
optical damping rate (net cooling rate) given by
\end{subequations}
\begin{subequations}
\begin{gather}
\frac{{1}}{{\Gamma}}=\frac{{1}}{{\Gamma}_{\mathrm{wk}}}+\frac{{1}}{{\Gamma
}_{\mathrm{str}}},\\
{\Gamma}_{\mathrm{wk}}=\frac{{4G}^{2}}{{\kappa}},\text{ \ }{\Gamma
}_{\mathrm{str}}={\kappa.}%
\end{gather}
Here ${\Gamma}_{\mathrm{opt}}^{\mathrm{wk}}$ and ${\Gamma}_{\mathrm{opt}%
}^{\mathrm{str}}$ represent the optical damping rate in the weak and strong
coupling regimes, respectively. Therefore, for the weak coupling case, to
obtain a high cooling rate, one expect a large ${G}^{2}/{\kappa}$; while in
the strong coupling regime, a large ${\kappa}$ leads to a high cooling rate.

On the other hand, large $G$ and large $\kappa$ result in higher quantum
cooling limit $n_{\mathrm{s}}^{\mathrm{(0)}}$ due to stronger quantum
backaction, as plotted in Fig. \ref{Fig2}(b). These trade-offs result in
optimal $\kappa$ and $G$ for the total cooling limit $n_{\mathrm{s}}$, which
can be approximately derived as%

\end{subequations}
\begin{subequations}
\begin{align}
{\kappa}_{\mathrm{opt}}  &  \simeq1.5{\omega_{\mathrm{m}}}(\frac
{n_{\mathrm{th}}}{Q_{\mathrm{m}}})^{\frac{1}{3}}{,}\label{ks}\\
{G}_{\mathrm{opt}}  &  \simeq0.9{\omega_{\mathrm{m}}}(\frac{n_{\mathrm{th}}%
}{Q_{\mathrm{m}}})^{\frac{1}{3}}, \label{Gs}%
\end{align}
where $Q_{\mathrm{m}}={\omega_{\mathrm{m}}/\gamma}$ denotes the mechanical
quality factor. It shows that ${G}_{\mathrm{opt}}\sim0.6{\kappa}%
_{\mathrm{opt}}$, indicating the intermediate coupling. The gray dotted
vertical and horizontal lines in Fig. \ref{Fig2}(c) denote $\kappa={\kappa
}_{\mathrm{opt}}$ and $G={G}_{\mathrm{opt}}$, which agree well with the
numerical results.

In Fig. \ref{Fig3} we further plot $n_{\mathrm{s}}^{\mathrm{(1)}}$,
$n_{\mathrm{s}}^{\mathrm{(0)}}$ and $n_{\mathrm{s}}$ for optimized $G$ and
${\kappa}$, along the horizontal and vertical lines in Fig. \ref{Fig2}(c),
respectively. It shows that classical cooling limit dominates for small
$\kappa$ and $G$, while quantum cooling limit becomes important as $\kappa$
and $G$ increase, which are precisely described by Eqs. (\ref{nscR}) and
(\ref{nsqR}).

With the optimal parameters given in Eqs. (\ref{ks}) and (\ref{Gs}), the
optimal cooling limit reads
\end{subequations}
\begin{equation}
n_{\mathrm{opt}}\simeq1.8(\frac{n_{\mathrm{th}}}{Q_{\mathrm{m}}})^{\frac{2}%
{3}}.
\end{equation}
In Fig. \ref{Fig3}(c) we plot $n_{\mathrm{opt}}$ as a function of the
environmental temperature $T$ for various $Q$-frequency products. It shows
that a high $Q$-frequency product allows for achieving a low phonon number at
a high temperature region. For ground-state cooling (or ground-state occupancy
probability $P>50\%$), it requires
\begin{equation}
Q_{\mathrm{m}}>2.4n_{\mathrm{th}}. \label{Qm}%
\end{equation}
For typical mechanical resonators, $\hbar{\omega_{\mathrm{m}}\ll
k}_{\mathrm{B}}T$, and the thermal phonon number is approximated as
$n_{\mathrm{th}}\simeq{k}_{\mathrm{B}}T/(\hbar{\omega_{\mathrm{m}}})$.
Therefore, the condition (\ref{Qm}) is equivalent to $Q_{\mathrm{m}}%
{\omega_{\mathrm{m}}>}2.4{k}_{\mathrm{B}}T/\hbar$. Starting from room
temperature ($T=300$ $\mathrm{K}$), the requirement for ground-state cooling
is expressed by the $Q$-frequency product
\begin{equation}
Q_{\mathrm{m}}{\nu>1.5}\times10^{13},
\end{equation}
where ${\nu=\omega_{\mathrm{m}}/}2\pi$ is the mechanical resonance frequency.

\begin{figure}[tb]
\centerline{\includegraphics[width=\columnwidth]{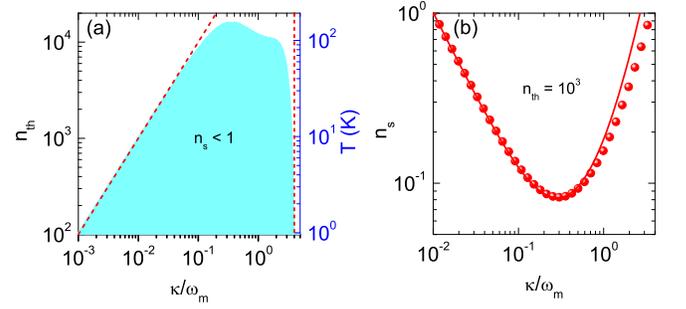}}
\caption{(color online) (a) Contour plot of the ground-state region
($n_{\mathrm{s}}<1$) as functions of ${\kappa}$ and $n_{\mathrm{th}}$. The
shaded region corresponds to $n_{\mathrm{s}}<1$. The red dashed lines denote
${\kappa=\gamma}n_{\mathrm{th}}$ (left) and ${\kappa=4\omega_{\mathrm{m}}}$
(right), respectively. (b) Cooling limit $n_{\mathrm{s}}$ as a function of
${\kappa}$ for $n_{\mathrm{th}}=10^{3}$. The red circles are the numerical
results and the red solid curve corresponds to the analytical results given by
Eqs. (\ref{nscR}) and (\ref{nsqR}). Other parameters: $\Delta^{\prime
}=-{\omega_{\mathrm{m}}}$, ${\gamma/\omega_{\mathrm{m}}=10}^{-5}$ and
${G=G}_{\mathrm{opt}}({\kappa})$ [given by Eq. (\ref{Goptk})]. }%
\label{Fig4}%
\end{figure}

In real experiments, there are restrictions on the cavity decay rate $\kappa$
and the coupling strength $G$. For example, many optical cavities have lower
bounds for $\kappa$ due to the limitation of fabrication and material
absorption. The coupling strength $G$ is related to the intracavity optical
field, while strong light field usually leads to material absorption and
heating. Therefore, it is important to take these constraints into
consideration. In the following we provide the parameter range for $\kappa$
and $G$ where ground-state cooling can be reached.

First we consider the requirement for the cavity decay rate $\kappa$. The
optimal coupling strength $G$ for a given ${\kappa}$ is obtained as
\begin{equation}
{G}_{\mathrm{opt}}({\kappa})\simeq(\frac{{\omega_{\mathrm{m}}^{2}\kappa\gamma
}n_{\mathrm{th}}}{2})^{\frac{1}{4}}. \label{Goptk}%
\end{equation}
Under this condition, ground-state cooling requires
\begin{equation}
{\gamma}n_{\mathrm{th}}<{\kappa<4\omega_{\mathrm{m}}}. \label{conk}%
\end{equation}
The left inequality reveals that the cavity decay rate should exceed the
thermal decoherence rate ${\Gamma}_{\mathrm{th}}={\gamma}n_{\mathrm{th}}$ to
suppress the environmental heating. The requirement can be re-expressed as
$Q_{\mathrm{m}}{\kappa>k}_{\mathrm{B}}T/\hbar$. At room temperature, it
yields
\begin{equation}
Q_{\mathrm{m}}{\kappa>6.2}\times10^{12}\text{.}%
\end{equation}
The right inequality in Eq. (\ref{conk}) shows that the resolved sideband
condition should be satisfied to reduce the quantum backaction heating. In
Fig. \ref{Fig4}(a) the exact numerical results for the ground-state region is
plotted, with the region boundary well described by Eq. (\ref{conk}). As an
example, we plot the cooling limit $n_{\mathrm{s}}$ as a function of ${\kappa
}$ for ${\gamma/\omega_{\mathrm{m}}=10}^{-5}$ and $n_{\mathrm{th}}=10^{3}$. In
this case $n_{\mathrm{s}}<1$ requires ${0.01}<{\kappa/\omega_{\mathrm{m}}<4}$.

\begin{figure}[tb]
\centerline{\includegraphics[width=\columnwidth]{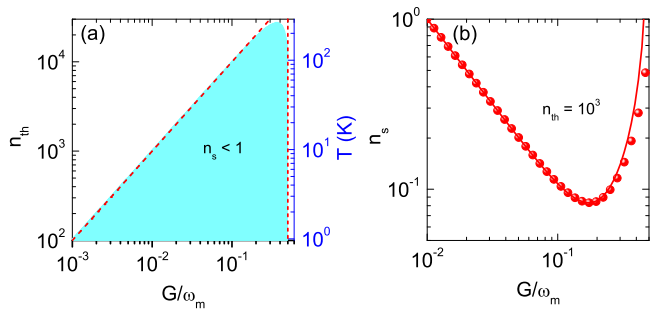}}
\caption{(color online) Same as Fig. \ref{Fig4} except that the horizontal
axes are ${G}$ and the value of the cavity decay rate is ${\kappa=\kappa
}_{\mathrm{opt}}({G})$ [given by Eq. (\ref{koptG})]. In (a), the red dashed
lines denote ${G=\gamma}n_{\mathrm{th}}$ (left) and ${G=0.5\omega_{\mathrm{m}%
}}$ (right), respectively.}%
\label{Fig5}%
\end{figure}

To obtain the requirement for the coupling strength $G$, we determine the
optimal cavity decay rate $\kappa$ for a given $G$ as
\begin{equation}
{\kappa}_{\mathrm{opt}}({G})\simeq2G. \label{koptG}%
\end{equation}
Then the requirement for ground-state cooling is given by
\begin{equation}
{\gamma}n_{\mathrm{th}}<{G<0.5\omega_{\mathrm{m}}}. \label{conG}%
\end{equation}
Clearly, the coupling strength should also exceed the thermal decoherence rate
to suppress the environmental heating, and room-temperature ground-state
cooling requires
\begin{equation}
Q_{\mathrm{m}}{G>6.2}\times10^{12}.
\end{equation}
The upper restriction ${G<0.5\omega_{\mathrm{m}}}$ is limited by the stability
condition given by Eq. (\ref{Stb2}). In Fig. \ref{Fig5}(a) the exact numerical
results for the ground-state region is plotted, with the region boundary well
described by Eq. (\ref{conG}). A example for $n_{\mathrm{th}}=10^{3}$ is shown
in Fig. \ref{Fig5}(b).

In summary, we have examined the backaction cooling of mesoscopic mechanical
resonators in the intermediate coupling regime. We develop a general framework
to describe the steady-state backaction cooling limits in the full parameter
range. We have analytically derived the optimal cooling limits and the optimal
parameters including the cavity decay rate and the optomechanical coupling
strength. In the resolved sideband regime, under the optimal detuning
$\Delta^{\prime}=-{\omega_{\mathrm{m}}}$, the optimal cavity decay rate and
the optimal coupling strength are derived as ${\kappa}_{\mathrm{opt}}%
\simeq1.5{\omega_{\mathrm{m}}}(n_{\mathrm{th}}/Q_{\mathrm{m}})^{1/3}${ and
}${G}_{\mathrm{opt}}\simeq0.9{\omega_{\mathrm{m}}}(n_{\mathrm{th}%
}/Q_{\mathrm{m}})^{1/3}$, with the lowest cooling limit being $n_{\mathrm{opt}%
}\simeq1.8(n_{\mathrm{th}}/Q_{\mathrm{m}})^{2/3}$. At the optimal point, the
requirement for ground-state cooling is $Q_{\mathrm{m}}>2.4n_{\mathrm{th}}$.
Starting from room temperature, ground-state cooling is achievable for
mechanical $Q$-frequency product $Q_{\mathrm{m}}{\nu>1.5}\times10^{13}$. For
practical optomechanical systems, the allowed parameter regions for
ground-state cooling are $\gamma n_{\mathrm{th}}<{\kappa<4\omega_{\mathrm{m}}%
}$ and $\gamma n_{\mathrm{th}}<{G<0.5\omega_{\mathrm{m}}}$. This provides a
guideline for achieving the lowest cooling limit towards room-temperature
ground-state cooling of mechanical resonators.

\begin{acknowledgments}
This work is supported by the 973 program (2013CB921904, 2013CB328704), NSFC
(11004003, 11222440, and 11121091), and RFDPH (20120001110068). Y.C.L is
supported by the Scholarship Award for Excellent Doctoral Students granted by
the Ministry of Education.
\end{acknowledgments}

\end{document}